\documentclass[runningheads,fleqn]{llncs}

\usepackage{booktabs}   %% For formal tables:
\usepackage{subcaption} %% For complex figures with subfigures/subcaptions
\usepackage{latexsym}
\usepackage{amssymb}
\usepackage{amsmath}
\usepackage{stmaryrd}
\usepackage[all]{xy}
\usepackage{graphicx}
\usepackage{hyperref}
\usepackage{multicol}
\usepackage{bm}

\usepackage{tikz}
\usetikzlibrary{calc,arrows,shadows,positioning}

\newcommand{\Log}{\omega}

\newcommand{\ignore}[1]{}

\newcommand{\arro}[1]{\xrightarrow{#1}} 
\newcommand{\hoo}{\hookrightarrow}

\newcommand{\comp}{\:|\:}

\newcommand{\swb}[2]{#1 {}_{\langle\!\langle #2}}
\newcommand{\swa}[2]{#1 {}_{\rangle\!\rangle #2}}

\newcommand{\h}{\mathit{h}}

\renewcommand{\k}{\lambda}
\renewcommand{\k}{\ell}
\newcommand{\init}{\mathsf{init}}
\newcommand{\final}{\mathsf{final}}
\newcommand{\rlh}{\rightleftharpoons}
\newcommand{\lh}{\leftharpoondown}
\newcommand{\rh}{\rightharpoonup}

\newcommand{\sql}{\mbox{$\lfloor\hspace{-.5ex}\lfloor$}}
\newcommand{\sqr}{\mbox{$\rfloor\hspace{-.4ex}\rfloor$}}

\usepackage{xcolor}

\newcommand{\red}[1]{{\color{red} #1}}

\newcommand{\blue}[1]{{\color{blue} #1}}

\newcommand{\ol}[1]{\overline{#1}}  % sequence of objects

\newcommand{\cT}{{\mathcal{T}}}

\newcommand{\nil}{[\:]}

\newcommand{\trace}{\mathit{tr}}
\newcommand{\prolog}{\mathit{log}}

\newcommand{\sender}{\mathit{sender}}
\newcommand{\parent}{\mathit{parent}}

\newcommand{\cons}{\!:\!}
\newcommand{\conc}{\mbox{$+$}}

\def \tuple#1{\langle #1 \rangle}

\long\def\comment#1{}

\newcommand{\cauder}{\textsf{CauDEr}}

\newcommand{\spawn}{\mathsf{spawn}}
\newcommand{\send}{\mathsf{send}}
\newcommand{\deliver}{\mathsf{deliver}}
\newcommand{\rec}{\mathsf{rec}}
\newcommand{\exit}{\mathsf{exit}}

\newcommand{\spawnm}{\mathsf{sp}}

\newcommand{\exitm}{\mathsf{e}}

\newcommand{\race}{\mathsf{race\_set}}
\newcommand{\rdep}{\mathsf{rdep}}
\newcommand{\variant}{\mathsf{variant}}

\newcommand{\actions}{\mathsf{actions}}

\newcommand{\grr}{\rightarrowtail}

\usepackage{accents}

\makeatletter
\providecommand{\leftsquigarrow}{%
  \mathrel{\mathpalette\reflect@squig\relax}%
}
\newcommand{\reflect@squig}[2]{%
  \reflectbox{$\m@th#1\rightsquigarrow$}%
}
\makeatother

\begin{document}

\title{%
%Combining State-Space Exploration and Causal-Consistent Reversible %Debugging
A Lightweight Approach to Computing Message Races 
%in Message-Passing Concurrent Programs 
with an Application to Causal-Consistent Reversible Debugging
%with Selective Receives%
%with an Application to Causal-Consistent Replay Debugging%
%and its 
%Application in Causal-Consistent Replay Debugging%
\thanks{
This work has been partially supported by grant PID2019-104735RB-C41
funded by MCIN/AEI/ 10.13039/501100011033, by the 
\emph{Generalitat Valenciana} under grant Prometeo/2019/098 
(DeepTrust), and by
    French ANR project DCore ANR-18-CE25-0007.
}
}
%\subtitle{\em\rm (submitted)}

\titlerunning{A Lightweight Approach to Computing Message Races}

\author{Juan Jos\'e Gonz\'alez-Abril  \and Germ\'an Vidal}
\authorrunning{J.J. Gonz\'alez-Abril  and G. Vidal}

\institute{
MiST, VRAIN, Universitat Polit\`ecnica de Val\`encia\\
\email{juagona6@vrain.upv.es~~gvidal@dsic.upv.es}
}

\maketitle

\begin{abstract}
	This paper presents a lightweight formalism (a trace) to model
	message-passing concurrent executions where some common
	common problems can be identified, 
	like lost or delayed messages, some forms of deadlock, etc.
	In particular, we consider (potential) message races that
	can be useful to analyze alternative executions. 
	We consider a particular application for our developments
	in the context of a causal-consistent reversible debugging 
	framework for Erlang programs.
\end{abstract}

%\keywords{concurrency, logging, causal-consistent
%  debugging}

%%%%%%%%%%%%%%%%%%%%%%%%%%%%%%%%%%%%%%%%%%%%%
\section{Introduction} \label{sec:intro}

Program debugging is generally a difficult task. When we
observe a misbehavior during a program execution 
(the \emph{symptom}), finding
the source of the error is often a challenging task.
This is particularly difficult in the context of 
concurrent and distributed software due to the
inherent nondeterminism of executions. As a consequence,
reproducing a faulty execution in a debugger is rather
difficult. 
Several debugging techniques have been developed in
order to overcome this problem. Typically, programs are
instrumented in order to generated a form of log (or trace)
of an execution. These logs can the be used to 
analyze what went wrong in the execution (as in, e.g.,
\emph{post-mortem} debugging) or they can be loaded
in a debugger in order to reproduce the faulty
execution (as in, e.g., \emph{record and replay}
debugging).

In this work, we consider a message-passing concurrent
language like Erlang \cite{erlang}. The language essentially
follows the actor model so that, at runtime, an application
consists of a number of processes that can only interact
through (asynchronous) message sending and 
receiving.\footnote{In practice, some Erlang built-in's 
involve shared-memory concurrency,
but we will not consider them in this work.}
Each process has a (private) mailbox, where messages are
stored until they are eventually consumed (or become
\emph{orphan} messages). 
Executions are typically nondeterministic because of
the order in which messages are delivered to processes.
Consider, for instance, two processes, $\mathsf{A}$ and
$\mathsf{B}$, that send messages $\mathsf{m_A}$
and $\mathsf{m_B}$, respectively, to
another process $\mathsf{C}$. If $\mathsf{A}$ and 
$\mathsf{B}$ are independent (i.e., their actions do
not depend on each other), then messages
$\mathsf{m_A}$ and $\mathsf{m_B}$ can reach
process $\mathsf{C}$ in any order, a so-called
\emph{message race}. Here, the order
in which this messages are delivered may determine
the rest of the execution (and, ultimately, the outcome
of the execution). 
Actually, we might have very unusual \emph{interleavings}
that do not happen during program testing but eventually
arise after software deployment, producing unexpected
and hard to find errors.

In order to improve software reliability, one can
explore all possible interleavings of a program, 
checking that no errors may arise in any of them. 
This is the core idea of techniques like 
\emph{stateless model checking} \cite{God97}
and 
\emph{reachability testing} \cite{LC06}.
Intuitively speaking, one can proceed as follows:
\begin{itemize}
\item First, a random execution of the program is considered.
\item Then, message races in this execution are
identified, i.e., situations like that of messages
$\mathsf{m_A}$ and $\mathsf{m_b}$ described above.
\item For each message race, alternative interleavings 
are considered. These alternatives are used to \emph{drive}
a new execution up to the point where messages are
delivered in a different order, then continuing 
nondeterministically as usual.
\item This process is repeated until all possible interleavings
of a program have been considered. In practice, 
dynamic partial order reduction (DPOR) techniques 
\cite{FG05} are typically considered to avoid 
exploring \emph{equivalent} executions once and again.
\end{itemize}
In this work, however, we consider a different application
of message races in the context of a
\emph{reversible debugging} 
framework for Erlang \cite{LPV19,LPV21}
and the associated tool \cauder\ \cite{cauder}.
Reversible debugging allows one to find the source
of bugs in a more natural way by exploring a faulty
execution from the observed misbehavior back to its
cause. Moreover, \cauder\ includes a \emph{replay}
mode where particular executions can be reproduced,
and explored back and forth, using execution logs.
In this context, the computation of 
message races can be useful to show the user 
those points in an execution where a different interleaving
is possible. In these cases, the user may decide to
abandon the execution that was being replayed and
explore a different one.

%
%%The paper is organized as follows. After some motivation 
%%in Section~\ref{sec:motivation}, we present the main
%%technical contributions in Section~\ref{sec:races}. Then,
%%Section~\ref{sec:applications} presents two applications for
%%our developments in the context of the language Erlang.
%Finally,  we present some related work (Section~\ref{sec:relwork}),
%and conclude (Section~\ref{sec:concl}).

%%%%%%%%%%%%%%%%%%%%%%%%%%%%%%%%%%%%%%%%%%%%%
\section{Message-Passing Concurrent Executions and Logs} \label{sec:motivation}

Let us start by considering the notion of  \emph{log} as defined
by Lanese et al \cite{LPV19,LPV21}, which has some similarities
with the \emph{SYN-sequences} of reachability testing \cite{LC06}.
Both formalisms represent a \emph{partial order} for the
executed actions, which contrasts with the interleavings
considered, e.g., in stateless model checking  \cite{God97}
and DPOR techniques \cite{FG05,AAJS17}.
To be precise, a log maps each process to
a sequence of the following actions:
\begin{itemize}
\item process spawning, denoted by $\spawn(p)$, where $p$
is the \emph{pid} (process identifier, which is unique) 
of the new process;
\item message sending, denoted by $\send(\ell)$, 
where $\ell$ is a (unique) message tag;
\item and message reception, denoted by $\rec(\ell)$, 
where $\ell$ is a message tag.
\end{itemize}
%In contrast to \cite{LC06}, \emph{synchronization pairs}
%are not explicitly considered but can easily be inferred 
%from send/receive actions with the same message tag.
%Note that $\spawn$ actions are needed because  runtime
%processes are not statically fixed; otherwise, we could skip
%these actions.
%
Execution logs represent quite a rough abstraction of an 
actual execution,  but they have enough information to make 
a message-passing  execution essentially
\emph{deterministic} \cite{LPV19,LPV21}.
This is why logs are used by the causal-consistent 
reversible debugger \cauder\ \cite{cauder} as part of an 
approach to record-and-replay debugging in Erlang. 

Unfortunately, logs do not contain enough information for
other purposes, like computing message races,
\emph{blocked} processes (a form of deadlock) 
or distinguishing \emph{lost} (i.e., not delivered)
messages from \emph{orphan} (i.e., delivered but not consumed)
messages. 
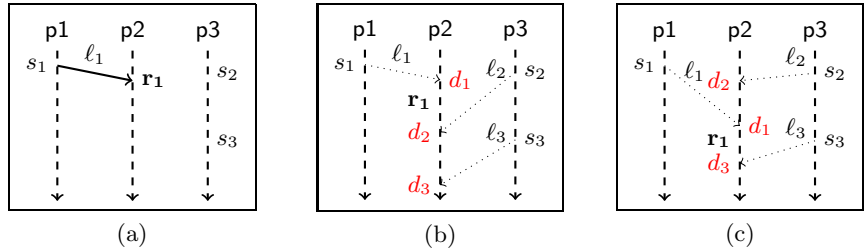
\begin{figure}[t]
\begin{minipage}{.3\linewidth}
\centering
\fbox{\begin{tikzpicture}
\draw[->,dashed,thick] (0,2) node[above]{\sf p1} -- (0,0);
\draw[->,dashed,thick] (1,2) node[above]{\sf p2} -- (1,0);
\draw[->,dashed,thick] (2,2) node[above]{\sf p3} -- (2,0);

\draw[->, thick] (0,1.8) node[left]{$s_1$} -- (1,1.6) node[midway,above]{$\ell_1$} node[right]{$\mathbf{r_1}$};

\node at (2,1.7) [right] {$s_2$};
\node at (2,0.8) [right] {$s_3$};
\end{tikzpicture}}\\[1ex]
(a)
\end{minipage}
~~
\begin{minipage}{.3\linewidth}
\centering
\fbox{\begin{tikzpicture}
\draw[->,dashed,thick] (0,2) node[above]{\sf p1} -- (0,0);
\draw[->,dashed,thick] (1,2) node[above]{\sf p2} -- (1,0);
\draw[->,dashed,thick] (2,2) node[above]{\sf p3} -- (2,0);

\draw[->, dotted] (0,1.8) node[left]{$s_1$} -- (1,1.6) node[midway,above]{$\ell_1$} node[right]{\red{$d_1$}};

\node at (1,1.3) [left] {$\mathbf{r_1}$};

\draw[->, dotted] (2,1.7) node[right]{$s_2$} -- (1,0.9) node[near start,above]{$\ell_2$} node[left]{\red{$d_2$}};

\draw[->, dotted] (2,0.8) node[right]{$s_3$} -- (1,0.2) node[near start,above]{$\ell_3$} node[left]{\red{$d_3$}};
\end{tikzpicture}}\\[1ex]
(b)
\end{minipage}
~
\begin{minipage}{.3\linewidth}
\centering
\fbox{\begin{tikzpicture}
\draw[->,dashed,thick] (0,2) node[above]{\sf p1} -- (0,0);
\draw[->,dashed,thick] (1,2) node[above]{\sf p2} -- (1,0);
\draw[->,dashed,thick] (2,2) node[above]{\sf p3} -- (2,0);

\draw[->, dotted] (0,1.8) node[left]{$s_1$} -- (1,1) node[pos=0.4,above]{$\ell_1$} node[right]{\red{$d_1$}};

\draw[->, dotted] (2,1.7) node[right]{$s_2$} -- (1,1.6) node[near start,above]{$\ell_2$} node[left]{\red{$d_2$}};

\node at (1,0.8) [left] {$\mathbf{r_1}$};

\draw[->, dotted] (2,0.8) node[right]{$s_3$} -- (1,0.5) node[near start,above]{$\ell_3$} node[left]{\red{$d_3$}};
\end{tikzpicture}}\\[1ex]
(c)
\end{minipage}
\caption{Some possible message-passing diagrams. 
We have three processes, identified  by  pids 
$\mathsf{p1}$, $\mathsf{p2}$
and $\mathsf{p3}$. Solid arrows denote the connection between
messages sent and received (similarly to the synchronization 
pairs of \cite{LC06}), while dotted arrows represent 
message delivery. Time, represented
by dashed lines, flows from top to bottom.} \label{fig:problema}
\end{figure}
Let us illustrate these issues with an example. Consider 
the following simple log:
\[
[\mathsf{p1}\mapsto \spawn(\mathsf{p2}),\spawn(\mathsf{p3}),\send(\ell_1);~~
\mathsf{p2}\mapsto  \rec(\ell_1);~~
\mathsf{p3}\mapsto \send(\ell_2),\send(\ell_3)]
\]
where $\mathsf{p1},\mathsf{p2},\mathsf{p3}$
are pids and $\ell_1,\ell_2,\ell_3$ are message tags.
Here, we can easily see that message $\ell_1$ has been
\emph{consumed} by process $\mathsf{p2}$ since the log
contains the following pair of elements: 
$\send(\ell_1)$ and $\rec(\ell_1$). This execution can be
represented by the message-passing diagram in
Figure~\ref{fig:problema}a, where $s_i$ denotes
a concurrent action  of the form $\send(\ell_i)$, $i=1,\ldots,3$,
and $\mathbf{r_1}$ denotes $\rec(\ell_1)$. 

Unfortunately, the information in the above log is not enough to 
identify message races. For this purpose, we need to
know (at least) the target of each message. For instance, we
can replace $\send(\ell)$ by $\send(\ell,p)$ in the logs,
where $p$ is the pid of the target process:
\[
\begin{array}{lll}
[ & \mathsf{p1}\mapsto \spawn(\mathsf{p2}),\spawn(\mathsf{p3}),\send(\ell_1,\mathsf{p2});\\
& \mathsf{p2}\mapsto  \rec(\ell_1);\\
& \mathsf{p3}\mapsto \send(\ell_2,\mathsf{p2}),\send(\ell_3,\mathsf{p2}) & ]
\end{array}
\]
Now, we can see that all three messages, $\ell_1$,
$\ell_2$, and $\ell_3$ are addressed to the same process,
$\mathsf{p2}$. However, we do not know when
these messages were \emph{delivered}. As a consequence, the
same log might represent both the message-passing
diagram in Figure~\ref{fig:problema}b and that in
Figure~\ref{fig:problema}c. Hence, we cannot know where
there is a message race for the receive $\mathbf{r_1}$ between
messages $\ell_1$ and $\ell_2$ 
(the case in Figure~\ref{fig:problema}b)
or between messages $\ell_1$ and $\ell_3$ 
(the case in Figure~\ref{fig:problema}c). 
Observe that message $\ell_2$ cannot race with $\ell_1$
in Figure~\ref{fig:problema}c since $\ell_2$
was delivered before $\ell_1$. This situation typically
denotes that the value of message $\ell_2$ does not
meet the constraints of receive $\mathbf{r_1}$ and,
thus, was ignored.

Now, we add explicit actions for message delivery;
namely, we add $\deliver(\ell)$ to denote the delivery
of message $\ell$, which is represented by $d_i$
in Figure~\ref{fig:problema}. Then, the log represented in 
Figure~\ref{fig:problema}b is as follows:
\[
\begin{array}{lll}
[ & \mathsf{p1}\mapsto \spawn(\mathsf{p2}),\spawn(\mathsf{p3}),\send(\ell_1,\mathsf{p2});\\
& \mathsf{p2}\mapsto  \red{\deliver(\ell_1)},\rec(\ell_1),\red{\deliver(\ell_2)},\red{\deliver(\ell_3)};\\
& \mathsf{p3}\mapsto \send(\ell_2,\mathsf{p2}),\send(\ell_3,\mathsf{p2}) & ]
\end{array}
\]
while that of Figure~\ref{fig:problema}c is as follows:
\[
\begin{array}{lll}
[ & \mathsf{p1}\mapsto \spawn(\mathsf{p2}),\spawn(\mathsf{p3}),\send(\ell_1,\mathsf{p2});\\
& \mathsf{p2}\mapsto  \red{\deliver(\ell_2)},\red{\deliver(\ell_1)},\rec(\ell_1),\red{\deliver(\ell_3)};\\
& \mathsf{p3}\mapsto \send(\ell_2,\mathsf{p2}),\send(\ell_3,\mathsf{p2}) & ]
\end{array}
\]
Finally, we also add explicit events for process termination,
$\exit$. For instance, the log represented by the diagram
in Figure~\ref{fig:problema}b could now be as follows:
\[ 
\label{eqn:log}
\begin{array}{lll}
[ & \mathsf{p1}\mapsto \spawn(\mathsf{p2}),\spawn(\mathsf{p3}),\send(\ell_1,\mathsf{p2}),\red{\exit};\\
& \mathsf{p2}\mapsto  {\deliver(\ell_1)},\rec(\ell_1),{\deliver(\ell_2)},{\deliver(\ell_3)};\\
& \mathsf{p3}\mapsto \send(\ell_2,\mathsf{p2}),\send(\ell_3,\mathsf{p2}),\red{\exit} & ]
\end{array}
\tag{*}
\]
In this way, 
we can now easily identify the following issues
from a log:
\begin{itemize}
\item A process $p$ is \emph{blocked} (a form of deadlock) if
its log does not end with $\exit$, since all processes are 
assumed to exit eventually in a normal execution.

\item A message $\ell$ is a \emph{lost message} 
whenever the log includes $\send(\ell,p)$ but it does
not include $\deliver(\ell)$. 

\item Finally, a message $\ell$ is an \emph{orphan message}
whenever the log includes $\deliver(\ell)$ but it does
not include $\rec(\ell)$.
\end{itemize}
In the following, we use the term \emph{trace} for the extended
logs (including the modified $\send$ as well as the
new $\deliver$ and $\exit$ actions), while we keep
the word \emph{log} for the original notion introduced
in \cite{LPV19}.

The computation of message races (from a trace)
will be shown in the next section. 
Nevertheless, let us clarify that we follow a \emph{lightweight}
approach to the computation of message races so that
they are only \emph{potential} races. 
For instance, given the trace (\ref{eqn:log}) above 
(represented in Figure~\ref{fig:problema}b), we would
determine that there is a (potential) message race between
$\ell_1$ and $\ell_2$ for $\mathbf{r_1}$. However, if
message $\ell_2$ does not meet the constraints of
receive $\mathsf{r_1}$, the race might be between
$\ell_1$ and $\ell_3$ instead. 
In the following, we would compute both possibilities
and leave the user (or the debugging tool) to determine
where a race is indeed an actual race or not. 
An experimental evaluation to determine the success ratio
of this simple strategy is planned.
For a more elaborated approach that computes 
actual message races we refer the
interested reader to \cite{Vid22submitted}.

%%% Has pasado la sección intermedia (donde se formalizan
%%% derivaciones, trazas, logs, y se demuestran cosas,
%%% tras el \end{document} !!

%%%%%%%%%%%%%%%%%%%%%%%%%%%%%%%%%%%%%%%%%%%%%%%%%%%%%%%%%%
\section{Message Races and Reversible Debugging}\label{sec:applications} 

In this section, we formalize an extension of the causal-consistent
reversible debugging framework for Erlang \cite{LPV19,LPV21}
in order to also compute message races, following
the ideas presented so far.

\emph{Causal-consistent reversible debugging} \cite{GLM14}
allows one to inspect the execution of a concurrent
program back and forth,
similarly to so-called \emph{time-travel} debugging. 
Reversible debugging can be useful to debug issues
easier by ``rewinding" a faulty execution back to the source of the 
problem (in contrast to traditional debuggers that usually require 
several runs of the program, possibly including breakpoints).
Reversibility is particularly challenging in the context of 
concurrent and distributed applications since exploring
backwards a forward computation in exactly the inverse order 
is often a poor strategy (e.g., because a huge number of actions can be
completely unrelated to the process of interest). 
Here, causal-consistent 
reversible debugging can greatly improve the situation.
Compared to traditional reversible debuggers, \emph{causal-consistent}
reversible debuggers allow the user to undo the steps of a
concurrent execution in any arbitrary order, as long as the steps
are \emph{causal-consistent}, i.e.,
no action is undone until all the actions that depend on it 
have already been undone. For example, one cannot undo the
spawning of a process until all the actions of this process have
been undone. 
Therefore, causal-consistent reversibility is essential to avoid 
exploring a large number of 
unrelated execution steps.

These notions have been adapted
to a message-passing concurrent language like Erlang 
in \cite{NPV16b,LNPV18jlamp} and materialized in the
\cauder\ debugger
\cite{LNPV18jlamp,LNPV18,GV21}.
The scheme has also been extended to consider 
\emph{replay} debugging in \cite{LPV19,LPV21}, thus having the
advantages of both (causal-consistent) 
reversible and record-and-replay
debugging.
A debugging session is typically driven by user \emph{requests}
like ``go forward until process $p$ is spawned" or 
``go backward up to the point where message $\ell$
was sent", etc.
In this context, adding the computation
of message races can be very useful for the user to
identify possible sources of nondeterminism; furthermore,
the generation of so-called 
\emph{race variants} will allow one to explore alternative execution
paths in a systematic way.
For instance, one could introduce new requests like ``go forward
until a deadlock is detected" or ``go forward until an
orphan message is produced", where all race variants are
systematically considered. 
If a problem is eventually found, the user 
can then use the current requests to explore (back and forth)
the buggy execution and try to identify the source of the error.
In this way, one could get the best of both worlds, 
systematic state-space exploration and causal-consistent
reversible debugging in a single tool.

%%%%%%%%%%%%%%%%%%%%%%%%%%%%%%%%%%%%
\subsection{A Tracing Semantics for Erlang}
\label{sec:tracing}

First, we formalize an appropriate semantics for a significant
subset of the language Erlang \cite{erlang} which can be used
to produce a trace of an execution as a side-effect.
Following \cite{NPV16b,LNPV18jlamp}, we consider a layered 
semantics: an \emph{expression semantics} defined on 
\emph{local states} (typically including an environment, 
an expression and a stack, see, e.g., \cite{GV21}) and a
\emph{system semantics}. 

We are not concerned with the details of the expression semantics
here. Let us only mention that, given a \emph{local state} $ls$, 
we denote by $ls \stackrel{z}{\to} ls'$
one evaluation step, where $ls'$ is the new local state and 
$z$ is a label that denotes the type of evaluation; for \emph{global} or
\emph{non-local} actions, 
this label includes the information which is required for the 
next layer of the semantics (the \emph{system} semantics) 
to perform the associated side-effects.
The local state typically 
contains an environment (a substitution),
an expression (to be evaluated) and a stack (see \cite{GV21} for
more details).   
Here, we consider the following labels:
\begin{itemize}
\item $\iota$: a local evaluation (e.g., a function call, an arithmetic operation, etc).
\item $\send(v,p)$: the evaluation requires sending a
message $v$ to process with pid $p$ as a side-effect.
\item $\rec(\kappa,cs)$: the evaluation requires \emph{receiving}
a message, % as a side-effect, where 
where $cs$ are the branches
of the receive statement
and
$\kappa$ is a sort of \emph{future} that will be bound (in the
system semantics) to the
expression in the selected branch.
\footnote{As 
in Erlang, we assume that a receive expression 
``\texttt{receive $cs$ end}''
has the form \texttt{receive $p_1$[when $g_1$] $ \to e_1$; \ldots; 
$p_n$[when $g_n$] $ \to e_n$ end} so that it looks for the oldest message
in the process' mailbox that matches some pattern $p_i$ and the
corresponding guard $g_i$ holds; then, it continues evaluating $e_i$.
When no message matches any pattern, execution is blocked
until a matching message reaches
the mailbox.} 

\item $\spawn(\kappa,ls_0)$: the  evaluation requires 
\emph{spawning} a new process as a side-effect. The new process
will start with the initial local state 
$ls_0$. In this case, variable $\kappa$ will be bound to the pid 
of the new process in the system semantics.
\end{itemize}
As mentioned before, we assume that processes are given
a \emph{pid} (\emph{p}rocess \emph{id}entifier) that uniquely
identifies them in a computation;\footnote{We do not specify how
these unique identifiers can be computed in a concurrent or distributed setting, but refer the interested reader 
to, e.g., \cite{Lam78}.} 
in the following, we often refer to ``process $p$" to mean 
``process with pid $p$"\!\!.
Analogously, every message
is wrapped with a \emph{tag} which is unique too, so that we can
distinguish messages even when they have the same value.
The domains of pids, $\cal P$, and tags,
$\cal L$, are disjoint.
In this work, we distinguish five (global) \emph{actions}:
\begin{itemize}
\item $\spawn(p)$, for process spawning; 
\item $\exit$, for process exit (termination);
\item $\send(\ell,p)$, for message sending; 
\item $\deliver(\ell)$, for message delivery; and
\item $\rec(\ell)$, for message reception.
\end{itemize}
Here, $p\in{\cal P}$ is a pid and $\ell\in{\cal L}$ is a message tag.
An \emph{event} is then a pair $p\cons a$, where $a$ is an action,
and $p$ is the pid of the process performing this action. We note
that message delivery is attributed to the target of the message.

Let us now consider an instrumented version of the system 
semantics, called \emph{tracing semantics}, where transitions
are labeled with the associated event.
A running system includes a number of processes, which
are defined  as follows:

\begin{definition}[process]
  A process is denoted by a configuration of the form 
  $\tuple{p,ls,q}$, where $p$ is
  the pid (process identifier) of the process, which is unique in a system,
  $ls$ is the local state
  and $q$ is the process mailbox (a list).
\end{definition}
A \emph{system} is then defined as a pair $\Gamma;\Pi$, where $\Gamma$
represents the network (sometimes called the \emph{global mailbox} 
\cite{LNPV18jlamp}
or the \emph{ether} \cite{SFB10}) and $\Pi$ is a pool of processes.

The network, $\Gamma$, is defined as a collection of queues, 
one per each  pair of (not necessarily different) processes. 
We use the notation $\Gamma[(p,p')\mapsto qs]$
either as a condition on $\Gamma$ or as a modification of $\Gamma$,
where $p,p'$ are pids and $qs$ is a (possibly empty) queue;
for simplicity, 
we assume that queues are initially empty for each pair of processes.
We use list notation for queues, where $\nil$ denotes the empty
queue and $m \cons qs$ denotes a queue with first element $m$
and tail $qs$; moreover, we let $qs\conc m$ denote the addition
of message $m$ at the end of queue $qs$.

The second component, $\Pi$, is denoted as 
$ \tuple{p_1,ls_1,q_1} ~\comp \cdots
\comp~\tuple{p_n,ls_n,q_n} $, where ``$\comp\!$'' represents an
associative and commutative operator. 
We often denote a \emph{system} as
$ \Gamma; \tuple{p,ls,q}\comp\Pi $ to point out that
$\tuple{p,ls,q}$ is an arbitrary process of the pool
(thanks to the fact that ``$\comp$'' is associative and
commutative).

The rules of the 
\emph{tracing semantics} are shown in 
Figure~\ref{fig:tracing-semantics}. A standard semantics
would be similar to the tracing semantics by removing transition
labels and \emph{unwrapping} messages. 
Let us briefly explain the rules:
%(we skip the labels of the transitions for now):
\begin{itemize}
\item Rule \textit{Exit} removes a process from the pool
when the local state
is \emph{final}, i.e., when the expression to be reduced is a data term.
Rule \textit{Local} just updates the local state of the selected 
process according to a transition of the expression semantics, while
rule \textit{Self} binds $\kappa$ to the pid of the current process.

\item Rule \textit{Spawn}  updates the local state, binds 
$\kappa$ to the pid of the new process
and adds a new initial process configuration
with local state $ls_0$ as a side-effect.

\item Rule \textit{Send} updates the local state and, moreover, adds a new message to the corresponding queue of the network as a side-effect.
Evaluating a send statement and adding the 
message to the network is considered an atomic operation.
In contrast to the standard semantics, a 
message $v$ is ``wrapped'' with a message identifier $\ell$
so that messages can be tracked in a computation.

\item Rule \textit{Deliver} nondeterministically (since $\Gamma$
might contain several queues with the same target process $p$) 
takes a message from the network 
and moves it to the corresponding process mailbox.

\item Finally, rule \textit{Receive} consumes a message from the process
mailbox using the auxiliary function \texttt{matchrec} that 
takes the local state $ls'$, the \emph{future} $\kappa$,
the branches of the receive expression $cs$, 
and the queue $q$. It then selects the oldest message
in $q$ that matches a branch in $cs$  (if any),
and returns a new local state $ls''$ (where $\kappa$ is 
 bound to the expression in the
selected branch),  a queue $q'$ (where the selected message
has been removed), and the label of the selected message
(which is needed to label the transition step).
%We also note that function $\mathsf{matchrec}$ must now
%first ``unwrap'' the message and, then, proceed as before;
%moreover, it also returns the identifier of the selected message
%so that the step can be labeled with the appropriate information.
\end{itemize}

\begin{figure}[t]
  $
  \begin{array}{r@{~~}c}
    (\mathit{Exit}) & {\displaystyle
      \frac{\mathit{final}(ls)}{\Gamma;\tuple{p,ls,q} 
        \comp \Pi \hoo_{\red{p:\exit}} \Gamma;\Pi}
      }\\[2ex]

    (\mathit{Local}) & {\displaystyle
      \frac{ls \arro{\iota} ls'}{\Gamma;\tuple{p,ls,q} 
        \comp \Pi \hoo_\epsilon \Gamma;\tuple{p,ls',q}\comp \Pi}
      }\\[3ex]

      (\mathit{Self}) & {\displaystyle
        \frac{ls \arro{\mathsf{self}(\kappa)} ls'}{\Gamma;\tuple{p,ls,q} 
          \comp \Pi 
           \hoo_\epsilon \Gamma;\tuple{p,ls'\{\kappa\mapsto p\},q} \comp \Pi}
      }\\[3ex]

      (\mathit{Spawn}) & {\displaystyle
        \frac{ls \arro{\mathsf{spawn}(\kappa,ls_0)}
          ls'~~\mbox{and}~~ \red{p'~\mbox{is a fresh pid}}}{\Gamma;\tuple{p,ls,q} 
          \comp \Pi 
           \hoo_{\red{p:\spawn(p')}} \Gamma;\tuple{p,ls'\{\kappa\mapsto p'\},q}\comp \tuple{p',ls_0,\nil} 
          \comp \Pi}
      }\\[3ex]      

    (\mathit{Send}) & {\displaystyle
      \frac{ls \arro{\mathsf{send}(v,p')} ls' ~\mbox{and}~\red{\k~
       \mbox{is a fresh symbol}}
      }{\Gamma[(p,p') \mapsto qs];\tuple{p,ls,q} 
        \comp \Pi                       
       \hoo_{\red{p:\send(\ell,p')}} \Gamma[(p,p') \mapsto qs\conc \{\k,v\}];\tuple{p,ls',q}\comp \Pi}
      }\\[3ex]

   (\mathit{Deliver}) & {\displaystyle
      \frac{}{\Gamma[(p',p)\mapsto \{\red{\ell},v\}\cons qs];\tuple{p,ls,q} 
        \comp \Pi \hoo_{\red{p:\deliver(\ell)}} \Gamma[(p',p) \mapsto qs];
        \tuple{p,ls,q\conc \{\ell,v\}}  \comp \Pi }
      }\\[3ex]

      (\mathit{Receive}) & {\displaystyle
        \frac{ls \arro{\mathsf{rec}(\kappa,cs)}
          ls' ~~\mbox{and}~~ \mathsf{matchrec}(ls',\kappa,cs,q) = 
          (ls'',q',\red{\ell})
          }
         {\Gamma; \tuple{p,ls,q} \comp \Pi
                     \hoo_{\red{p:\rec(\ell)}}  \Gamma;\tuple{p,ls'',q'}\comp \Pi}
      }
  \end{array}
  $
\caption{Tracing semantics} \label{fig:tracing-semantics}
\end{figure}

Observe that message sending is split into three different actions:
send (the message is stored in the network), delivery 
(the message is moved from the network to the mailbox of
the target process) and 
receive (the message is consumed and removed from the
mailbox). This contrasts with the semantics in
\cite{LPV19,LPV21} where message delivery is abstracted away 
(since the simpler notion of log was enough for defining a 
\emph{replay semantics}). 

Proving that the tracing semantics is a conservative extension of
the standard semantics is straightforward, since it is 
essentially equivalent to the system semantics in \cite{LNPV18jlamp}
with the addition of message tags and transition labels.

Note that the tracing semantics has two main sources of 
nondeterminism: selecting a process to apply a reduction rule,
and selecting the message to be delivered from the network 
(rule \textit{Deliver}).
Regarding the first point, one can for instance implement a 
\emph{round-robin} algorithm that performs a maximum 
number of transitions, then moves to another process, etc.
As for the selection of a message to be delivered, there are
several possible strategies. For instance, the \cauder\ debugger 
\cite{LNPV18jlamp,LNPV18,GV21} implements both a user-driven strategy 
(where the user selects one of the available messages) as well 
as a random selection. 
Another possibility would be implementing a strategy called
\emph{instant-delivery}, where sent messages are immediately
stored in the mailbox of the target process (this is actually the
common strategy in Erlang runtime 
environments).\footnote{Instant-delivery 
is the default strategy in the Erlang model checker 
Concuerror \cite{CGS13}.} This strategy can be
formalized in our setting by requiring rules
\textit{Send} and \textit{Deliver} to be applied always 
consecutively (and atomically) in a derivation. 

Given systems $\alpha_0,\alpha_n$, 
we call $\alpha_0 \hoo^\ast \alpha_n$, which is a shorthand for
$\alpha_0 \hoo_{p_1:r_1} \ldots \hoo_{p_{n-1}:r_{n-1}} \alpha_{n}$, $n\geq 0$, a
\emph{derivation}.  One-step
derivations are simply called \emph{transitions}. We use
$\delta,\delta',\delta_1,\ldots$ to denote derivations and 
$t, t', t_1,\ldots$ for transitions. 
A system $\alpha$ is said \emph{initial} if it has the form
$\nil;\tuple{p,ls,\nil}$, where $p$ is the pid of some initial
process and $ls$ is an initial local state containing 
the expression to be evaluated.
In the following, we assume that all derivations start with
an initial system.

The notion of trace is now formalized as follows:

\begin{definition}[trace] \label{def:trace}
  A \emph{trace} is a mapping from pids to sequences of actions.
  Given a trace $\tau$, we let $\tau(p)$ denote 
  the sequence of actions associated to process $p$ 
  in $\tau$. Also, $\tau[p\mapsto as]$ denotes that
  $\tau$ is an arbitrary trace such that $\tau(p) = as$;
  we use this notation either as a condition on $\tau$ or 
  as a modification of $\tau$.
  
  We  say that an event $p\cons a$ occurs in a trace 
  $\tau[p\mapsto as]$ if action $a$ occurs in sequence $as$.
  Moreover, we say that event $p_1\cons a_1$ \emph{precedes}
  event $ p_2\cons a_2$ in $\tau$, in symbols
 $p_1\cons a_1\prec_\tau p_2\cons a_2$,  
 if $p_1=p_2$, $\tau(p_1) = as$,
  and $a_1$ precedes $a_2$ in $as$; otherwise, the (partial) 
  relation is not defined. 
  Two traces, $\tau$ and $\tau'$, are \emph{equal},
  %in symbols $\tau\doteq\tau'$, 
  if they are identical up to  renaming of pids and tags. 

  Let $\delta$ be a derivation of the form 
  $\alpha_0 \hoo_{e_1} \alpha_1
  \ldots \hoo_{e_n} \alpha_{n+1}$, and let 
  $e'_1,\ldots,e'_m$, $m\leq n$, 
  be the subset of its non-null labels. 
  Then, we say that the sequence $e'_1,\ldots,e'_m$ is the
 \emph{trace} of the derivation $\delta$, in symbols 
 $\cT(\delta)$.
\end{definition}
Note that traces represent a \emph{partial order} where only the
order of actions within a process matters. This contrasts with
the notion of \emph{interleaving}
considered, e.g., in stateless model checking  
\cite{God97} and DPOR techniques \cite{FG05,AAJS17}, which 
could be seen as a particular linearisation of a trace.

\subsection{Adding the Computation of Message Races to \cauder}
\label{sec:cauder-races}

In practice, 
traces can be generated either by implementing an instrumented
interpreter (following the tracing semantics)
or by instrumenting a program so that its execution
produces a trace as a side-effect (see \cite{GV21lopstr}).
Currently, the publicly available debugger
\cauder\ \cite{cauder} consists of
the following two components \cite{LPV19,LPV21}:
\begin{itemize}
\item A \emph{tracer}, that instruments a source program so that
its execution (in the standard environment) produces a log of
the computation as a side-effect. 
\item A \emph{reversible} debugger, that is able to replay a particular
execution given a program and the log of an execution. The user can
explore the execution back and forth using \emph{requests}, e.g.,
``go forward until the sending of message $\ell$",
``go backward up to the point where process $p$ is spawned", etc.
\end{itemize}
In this context, we have first replaced the \emph{logging} semantics
of \cite{LPV19,LPV21} with the tracing semantics shown in
Figure~\ref{fig:tracing-semantics}, so that we  produce a trace
instead of a (simpler) log. 
Thanks to this change, message races can now be computed
and, morever, 
a log can still be extracted from the trace anyway (since it is
a simplification of a trace) in order 
to drive the replay of an execution.

\begin{definition}[log \cite{LPV19,LPV21}]
	A \emph{log} is a mapping from pids to sequences of
	simple actions of the form $\spawn(p)$, $\send(\ell)$
	and $\rec(\ell)$, where $p$ is a pid and $\ell$ is a 
	message tag. 
	We use the same notation conventions as for traces.
	
	Given a trace $\tau$, we let $\prolog(\tau)$ be the log
	obtained from $\tau$ by removing message delivery and
	exit actions, as well as replacing every action of the 
	form $\send(\ell,p)$ by $\send(\ell)$. 
\end{definition}
As mentioned before, our traces represent a partial order on the
global actions of a message-passing concurrent program. 
This partial order can be
formalized using the well-known \emph{happened-before} relation
\cite{Lam78}:

\begin{definition}[happened-before, independence]
\label{def:happened-before}
Let $\tau$ be a  trace including events
  $e_1 = (p_1\cons a_1)$ and $e_2 = (p_2\cons a_2)$, $e_1,e_2\in\tau$.
  We say that $e_1$ \emph{happened before} $e_2$, in symbols 
  $e_1\leadsto_\tau e_2$, 
%  (or simply $e_1\leadsto e_2$ when the
%  trace is clear from the context), 
  if  
  %$e_1 \prec_\tau e_2$ and 
  one of the following conditions hold:\footnote{We use
  ``$\_$'' as a placeholder to denote an arbitrary value. 
%  Note that the second argument of
%   $\send$ and $\deliver$ is not used. 
   %The relevance of these
  %values will be clear in the next section.
  }
  \begin{enumerate}
  \item $p_1 = p_2$, $a_1\neq\deliver(\_)$, $a_2\neq\deliver(\_)$,
  and $e_1\prec_\tau e_2$; 

  %\item $p_1 = p_2$, $a_1 = \deliver(\k)$ and $a_2 = \rec(\k)$;

  \item $p_1 = p_2$, $a_1 = \deliver(\k)$, $a_2 = \deliver(\k')$, 
  $\k\neq\k'$, and $e_1\prec_\tau e_2$; %, and $e_1\prec_\tau e_2$;

  \item $a_1 = \spawn(p_2)$; %, $p_1\neq p_2$;

  \item $a_1=\send(\k,\_)$ and $a_2=\deliver(\k)$;

  \item $a_1=\deliver(\k)$ and $a_2=\rec(\k)$;
  
  \item $p_1=p_2$ and $a_2=\exit$. 
  \end{enumerate}
  %As is common, we denote by $\leadsto^+_\tau$ the transitive
  %closure of $\leadsto_\tau$.  
  Moreover, if $e_1\leadsto_\tau e_2$ and $e_2\leadsto_\tau e_3$, 
  then $e_1\leadsto_\tau e_3$ (transitivity). 
  Two events $e_1$ and $e_2$ are \emph{independent} in 
  $\tau$ if $e_1 \not\leadsto_\tau e_2$ 
  and $e_2\not\leadsto_\tau e_1$.\footnote{Note that the meaning
  of $e_1\not\leadsto_\tau e_2$ is ``$e_1 \leadsto_\tau e_2$ 
  is not true".} 
\end{definition}
Intuitively, our definition for the happened-before relation 
can be explained as follows:
(1) the actions of a given process which are not message deliveries
follow a strict order imposed by the program code;
(2) the order of message deliveries only matters within the same process;
(3) the spawning of a process happens before 
all the actions of this process;
(4) the sending of a message happens before its delivery;
(5) message delivery happens before it is consumed
by a receive statement;
and (6) all actions of a process must precede its exit.

Intuitively speaking, we have a message race whenever there is a 
receive statement that can nondeterministically consume 
different messages, depending on the considered derivation 
(interleaving).
A race set collects all such alternative messages (if any).
Observe that race sets are defined on traces, i.e., message races
do not depend on a particular derivation but on the class of 
derivations represented by a given trace.

\begin{definition}[race set] \label{def:race-set}
	Let $\tau$ be a  well-defined trace with
	$e_d = (p\cons\deliver(\ell))\in\tau$ and
	$e_r =(p\cons\rec(\ell))\in\tau$. 
	Consider a message $\ell'\neq\ell$, with sending
	and deliver events
	$e'_s =(p'\cons\send(\ell',p))\in\tau$ and 
	$e'_d =(p\cons\deliver(\ell'))\in\tau$, respectively. 
	We say that messages $\ell$ and $\ell'$ race for $e_r$ in $\tau$
	if  $e'_d\not\prec_\tau e_d$ and $e_d\not\leadsto_\tau e'_s$.
	
	We let $\race_\tau(e_r)$ denote a set with a list of message
	tags $[\ell_1,\ldots,\ell_n]$ for each process $p^*$ (possibly 
	equal to $p$) with at least one racing message, 
	ordered by the corresponding sending actions,
	i.e., $p^*\cons\send(\ell_1,p)\prec_\tau\ldots\prec_\tau p^*
	\cons\send(\ell_n,p)$.
	
	For convenience, 
	we also let $[[\race_\tau(e_r)]]$ denote the set of all message
	tags in  $\race_\tau(e_r)$, i.e., the union of all
	messages in the computed lists.
\end{definition}
Intuitively speaking, the definition above requires the following
conditions for messages $\ell$ and $\ell'$ to race for the 
receive statement $e_r$:
\begin{enumerate}
\item The target of both messages is the same ($p$).

\item The delivery of message $\ell'$ does not
precede the delivery of message $\ell$, because that would
point out either that $\ell'$ has been consumed by another
receive statement or that it does not match the constraints
of the receive statement $e_r$ (otherwise, $e_r$ would have
consumed message $\ell'$ instead of message $\ell$).
This is the situation for $\ell_2$ in 
the diagram shown in Figure~\ref{fig:problema}c.

\item Finally, we check that the delivery of the
consumed message $\ell$ ($e_d$) does not happen before
the sending of message $\ell'$ ($e'_s$). The reason is that 
$e_d\leadsto_\tau e'_s$ would prevent the delivery of message 
$\ell'$ ($e'_d$) to happen \emph{before} the delivery of message
$\ell$ ($e_d$) in any  derivation, thus 
there would be no way message $\ell'$
could be consumed by receive $e_r$.
\end{enumerate}
When there are several racing messages from the same process,
they are grouped into a list that follows the order of the
corresponding sending events. 
Note that the lists of messages actually represent \emph{potential}
races since one should still check that they match the constraints
of the receive statement (see \cite{Vid22submitted} 
for a different approach
that computes  \emph{actual} message races). 

A so-called \emph{race variant} can easily be computed
from the trace and the messages in the race set as follows:
\begin{itemize}
\item Let $\tau$ be a trace including the
receive event $e_r = (p\cons \rec(\ell))$. Let $\ell'$ the
considered message in $\race_\tau(e_r)$. 

\item Now, we compute a new trace by 
removing from $\tau$ all events $e$ such
$e_r \leadsto_\tau e$ and, then, replacing $e_r$
by $\rec(\ell)$. Let $\tau'$ be the resulting trace.

\item Then, the $\prolog(\tau)$ can be used to
replay a new execution where the receive statement
associated to $e_r$ will consume message $\ell'$ instead
of $\ell$ (assuming it meets the constraints of the
receive).
\end{itemize}
When exploring race variants, one can
try receiving these messages in order
until a matching one is found.
Here, we are assuming that
we do not want to consider traces with \emph{delayed} messages.
Otherwise, one should consider not only the first matching
message but all of them.
Observe that we do not need to check that message $\ell$
indeed matches the constraints of the receive $e_r$ since 
the considered trace has been obtained from
an actual execution. 

%Now, within the debugger, 
%when the execution reaches a receive statement, the
%debugger can show a list of (potential) messages
%that race with the received one. The user can then
%explore alternative execution paths. 
%%

\newcommand{\nxt}{\mathsf{next}}
\newcommand{\admi}{\mathsf{admissible}}

\begin{figure}[p]
  $
  \begin{array}{c}
   \multicolumn{1}{l}{(\mathit{Exit})} \\[-2ex]
   {\displaystyle
      \frac{ls \not\arro{} ~~\mbox{and}~~
      \mathit{final}(ls)}{\red{\Log};\Gamma;\tuple{p,\blue{h},ls,q} 
        \comp \Pi \rh_{p:\exit} \red{\Log};\Gamma;
        \tuple{p,\blue{\exit(ls,q)\cons h},\exit,q} 
        \comp \Pi}
      }\\[2ex]          

    \multicolumn{1}{l}{(\mathit{Local})} \\[-2ex]
     {\displaystyle
      \frac{ls \arro{\tau} ls'}{\red{\Log};\Gamma;\tuple{p,\blue{h},ls,q} 
        \comp \Pi \rh_\epsilon %_{p:\epsilon,\{s\}} 
        \red{\Log};\Gamma;\tuple{p,\blue{\tau(ls)\cons h},ls',q}\comp \Pi}
      }\\[2ex]

          \multicolumn{1}{l}{(\mathit{Self})} \\[-2ex]
       {\displaystyle
        \frac{ls \arro{\mathsf{self}(\kappa)} ls'}{\red{\Log};\Gamma;
        \tuple{p,\blue{h},ls,q} 
          \comp \Pi 
           \rh_\epsilon %{p:\epsilon,\{s\}}
           \red{\Log};\Gamma;\tuple{p,\blue{\mathsf{self}(ls)\cons h},
           ls'\{\kappa\mapsto p\},q} \comp \Pi}
      }\\[2ex]

      \multicolumn{1}{l}{(\mathit{Spawn})} \\[-2ex]
       {\displaystyle
        \frac{ls \arro{\mathsf{spawn}(\kappa,ls_0)}
          ls'~~\mbox{and}~~ \red{\nxt_p(\Log) = (p',\Log')}
          %p'~\mbox{is a fresh pid}
          }{%\begin{array}{l}
          \red{\Log};\Gamma;\tuple{p,\blue{h},ls,q} 
          \comp \Pi %\\
          %\hspace{4ex} 
           \rh_{p:\spawn(p')} %_{\red{p:\spawn(p'),\{s,sp_p'\}}} 
           \red{\Log'};\Gamma;
           \tuple{p,\blue{\spawn(ls,p')\cons h},
           ls'\{\kappa\mapsto p'\},q}\comp \tuple{p',ls_0,\nil} 
          \comp \Pi
          %\end{array}
          }
      }\\[3ex]

    \multicolumn{1}{l}{(\mathit{Send})} \\[-2ex]
     {\displaystyle
      \frac{ls \arro{\mathsf{send}(p',v)} ls'
      ~\mbox{and}~\red{\nxt_p(\Log) = (\ell,\Log')}
      }{\begin{array}{l}
      \red{\Log};\Gamma[(p,p') \mapsto qs];\tuple{p,\blue{h},ls,q} 
        \comp \Pi\\                       
       \hspace{4ex} \rh_{p:\send(\ell,p')} 
       %{\red{p:\send(\ell),\{s,\ell^\Uparrow\}}} 
       \red{\Log'};\Gamma[(p,p') \mapsto qs\conc \{v,\k\}];
       \tuple{p,\blue{\send(ls,p',\{v,\k\})\cons h},ls',q}\comp \Pi
      \end{array}
       }
      }\\[4ex]

   \multicolumn{1}{l}{(\mathit{Deliver})} \\
    {\displaystyle
      \frac{
      \red{\admi_p(\Log,q) = \ell}
      %\red{\nxt(p,\omega) = (\ell,\omega')}
      }{%\begin{array}{l}
      \red{\Log};\Gamma[(p',p)\mapsto \red{\{v,\ell\}}\cons vs];
      \tuple{p,\blue{h},ls,q} 
        \comp \Pi %\\
        %\hspace{4ex} 
        \rh_{p:\deliver(\ell)} %{\red{p:\deliver(\ell),\{s,\ell^=\}}} 
        \red{\Log};\Gamma[(p',p) \mapsto vs];
        \tuple{p,\blue{h},ls,q\conc \red{\{v,\ell\}}}  \comp \Pi 
        %\end{array}
        }
      }
      \\[3ex]

      \multicolumn{1}{l}{(\mathit{Receive})} \\
       {\displaystyle
        \frac{ls \arro{\mathsf{rec}(\kappa,cs)}
          ls' ~~ \mathsf{matchrec}(ls',\kappa,cs,q) = 
          (ls'',q',\ell,\red{v},\red{i}) ~~\mbox{and}~~ 
          \red{\nxt_p(\Log) = (\ell,\Log')}
          }
         {\red{\Log};\Gamma; \tuple{p,\blue{h},ls,q} \comp \Pi
                     \rh_{p:\rec(\ell)} % _{\red{p:\rec(\ell),\{s,\ell^\Downarrow\}}}  
                     \red{\Log'};
                     \Gamma;
                     \tuple{p,\blue{\rec(ls,\ell,v,i)\cons h},ls'',q'}\comp \Pi}
      }\\
      
%      
%   (\mathit{Skip}) & {\displaystyle
%      \frac{\red{\nxt(p,\omega) = (\ell,\omega')}}{\Gamma[(p',p)\mapsto \red{\{v,\ell\}}\cons vs];
%      \tuple{p,\red{\omega},\blue{h},ls,q} 
%        \comp \Pi \hoo_{\red{p:\mathsf{skip}(\ell)}} \Gamma[(p',p) \mapsto vs];
%        \tuple{p,\red{\omega'},\blue{h},ls,q}  \comp \Pi }
%      }
  \end{array}
  $
\caption{Forward reversible (replay) semantics} \label{fig:forward}
\end{figure}

\begin{figure}[p]
  $
  \begin{array}{lcl}
      (\ol{\mathit{Exit}}) & 
      \red{\Log};\Gamma;
        \tuple{p,\blue{\exit(ls,q)\cons h},\exit,q'} 
        \comp \Pi
        \lh_{p:\exit}
        \red{\Log};\Gamma;\tuple{p,\blue{h},ls,q} \comp \Pi        
      \\[2ex]

    (\ol{\mathit{Local}}) &
	\red{\Log};\Gamma;\tuple{p,
        \blue{\tau(ls)\cons h},ls',q}\comp \Pi
        \lh_\epsilon %{p:\epsilon,\{s\}}  
    \red{\Log};\Gamma;\tuple{p,\blue{h},ls,q} \comp \Pi
    \\[2ex]
    
          (\ol{\mathit{Self}}) &
      \red{\Log};\Gamma;\tuple{p,\blue{\mathsf{self}(ls)\cons h},
           ls',q} \comp \Pi
           \lh_\epsilon %%{p:\epsilon,\{s\}}
         \red{\Log};\Gamma;\tuple{p,\blue{h},ls,q} \comp \Pi  
      \\[1ex]

      (\ol{\mathit{Spawn}}) &
	      \red{\Log[p\mapsto as]};\Gamma;
           \tuple{p,\blue{\spawn(ls,p')\cons h},ls',q}\comp \tuple{p',ls_0,\nil} 
          \comp \Pi \hspace{20ex}\\
         &  \hspace{20ex}\lh_{p:\spawn(p')} %{\red{p:\spawn(p'),\{s,sp_p'\}}}
          \red{\Log[p\mapsto\spawn(p')\cons as]};
          \Gamma;\tuple{p,\blue{h},ls,q} 
          \comp \Pi 
      \\[2ex]
      
    (\ol{\mathit{Send}}) &
      \red{\Log[p\mapsto as]};\Gamma[(p,p') \mapsto qs\conc \{v,\k\}];
      \tuple{p,\blue{\send(ls,p',\{v,\k\})\cons h},ls',q}\comp \Pi
      \\
      & \hspace{10ex} 
      \lh_{p:\send(\ell,p')} %_{\red{p:\send(\ell),\{s,\ell^\Uparrow\}}}
     \red{\Log[p\mapsto\send(\ell)\cons as]};
     \Gamma[(p,p') \mapsto qs];
     \tuple{p,\blue{h},ls,q} \comp \Pi
      \\[1ex]

   (\ol{\mathit{Deliver}}) &
    \red{\Log};\Gamma[(p',p) \mapsto vs];
        \tuple{p,\blue{h},ls,q\conc \red{\{v,\ell\}}}  \comp \Pi 
        \hspace{20ex}\\
      & \hspace{20ex}  \lh_{p:\deliver(\ell)} %  {\red{p:\deliver(\ell),\{s,\ell^=\}}}
      \red{\Log};\Gamma[(p',p)\mapsto \red{\{v,\ell\}}\cons vs];
      \tuple{p,\blue{h},ls,q} 
        \comp \Pi 
        \\[1ex]
  	  %\multicolumn{1}{r}{\mbox{where}~\rec(\_,\ell,\_,\_)\not\in h}

      (\ol{\mathit{Receive}}) &
       \red{\Log[p\mapsto as]};\Gamma;
       \tuple{p,\blue{\rec(ls,\ell,v,i)\cons h},ls',q'}\comp \Pi
       \hspace{20ex} \\
       % & \mbox{where}\\
       & \hspace{20ex}
      \lh_{p:\rec(\ell)}  %{\red{p:\rec(\ell),\{s,\ell^\Downarrow\}}}
      \red{\Log[p\mapsto\rec(\ell)\cons as]};
      \Gamma; \tuple{p,\blue{h},ls,q} \comp \Pi
      %& \red{\mathsf{put}(q',i,\{v,\ell\}) = q}
	  \\
	  & \multicolumn{1}{r}{\mbox{where}~\red{\mathsf{put}(q',i,\{v,\ell\}) = q}}
      \\
  \end{array}
  $
\caption{Backward reversible semantics} \label{fig:backward}
\end{figure}

For this purpose, the replay reversible semantics has
also been extended.
In the (extended) reversible semantics, 
we have two transition relations, $\rh$ and $\lh$, that
represent the forward (replay) semantics and the backward
semantics, respectively. 
The rules, shown in Figures~\ref{fig:forward} 
and \ref{fig:backward}, are similar to those of the
tracing semantics shown in Figure~\ref{fig:tracing-semantics}.
The main difference is that, now, there are two new components:
\begin{itemize}
\item On the one hand, we add a log $\Log$ to a system configuration,
which will be used in rules \emph{Send}, \emph{Receive},
\emph{Spawn} and \emph{Deliver} to \emph{drive} the execution.
Here, we use the auxiliary function $\nxt_p$ to return either
the information from the next action of process $p$ in the log (and 
delete this action) or a fresh identifier if the log is already empty
(in order to allow the user to continue exploring an execution 
when the log represents only a prefix).
Therefore, the rules can be used either in \emph{replay} mode
or in \emph{user-driven} mode, similarly to the semantics presented
in \cite{GV21}.
We also use the auxiliary function $\admi$ to determine the next message
that can be delivered according to the current values in the log
and the process mailbox.

\item On the other hand, each process is now augmented with
a \emph{history}, i.e., a list of terms with enough information to
undo each foward step.   Observe that in rule \emph{Receive} we
store a term of the form $\rec(ls,\ell,v,i)$. Here, the position $i$
in the process mailbox is required in order to guarantee
reversibility, since the message consumed by a receive
statement needs
not be the first one in the queue (recall that a receive statement
consumes the oldest message \emph{that matches the
constraints} of the receive statement). This contrasts with the
approach in \cite{LNPV18jlamp} where the complete mailbox was
stored. An advantage of our approach is that we can have
a more general notion of \emph{concurrent} actions than 
\cite{LNPV18jlamp}, which was unnecessarily restrictive because
of the decision of storing the complete mailbox. To be precise,
now, a message delivery and a message reception commute
(i.e., are independent) when the considered messages are different. 
In \cite{LNPV18jlamp}, message reception and message delivery
never commute.
\end{itemize}
The reversible semantics is then given by the
relation $\rlh$, which denotes the union of the forward and backward
relations, i.e., $\rlh \:= (\rh \cup\lh)$. An essential feature of these
transition relations is that they are \emph{causal-consistent}.
Loosely speaking, it means that no action can be undone using
$\lh$ until
all the actions that depend on this action have been undone, and that
no action can be performed with $\rh$ (in replay mode, i.e., when
the log is not empty) until all the actions that happened
before this action have been performed. 

While the nondeterministic relation $\rlh$ models the legal steps 
in the reversible debugger, a so-called controlled semantics 
can be defined on top of it. This \emph{controlled} semantics
is used to drive the exploration of a given execution using
\emph{requests} of the form
``go forward until the sending of message $\ell$'' (in replay mode)
or
``go backward up to the point where process $p$ is spawned'', etc.

More details on this approach can be found in
\cite{LNPV18jlamp,LPV19,LPV21}. We claim that the framework
of \cite{LPV19,LPV21} can be extended using the rules in
Figures~\ref{fig:forward} and \ref{fig:backward}, and that
all properties proved there remain true (in particular, the
causal consistency of the reversible semantics). 
The complete formalization
of this extension is the subject of ongoing work.

%%%%%%%%%%%%%%%%%%%%%%%%%%%%%%%%%%%%%%%%%%%%%%%%%%%%%%%%%%
\section{Related Work}\label{sec:relwork}

This work stems from the idea of improving causal-consistent
reversible replay debugging \cite{LPV19,LPV21} with the computation of 
message races, since this information might be useful for the user 
in order to explore alternative execution paths. Causal-consistent
replay debugging introduces a \emph{logging} semantics that
produces a log of an execution. This log has many similarities 
with the SYN-sequences of reachability testing \cite{LC06}.
Both formalisms represent a partial order with the actions performed
by a number of processes running concurrently (they basically
denote a \emph{Mazurkiewicz trace} \cite{Maz86}).
SYN-sequences
are then used to define a systematic testing algorithm based on
the notions of message race and race variant. In contrast to
\cite{LC06}, we consider both traces and logs, and our traces
are tailored to a language with selective receives by
explicitly distinguishing message delivery and reception,
which is not done in \cite{LC06}.

Both reachability testing and our approach share many similarities
with so-called \emph{stateless model checking}  \cite{God97}.
The main difference, though, is that stateless model checking works 
with interleavings.
Then, since many interleavings may boil down to the same
Mazurkiewicz trace, \emph{dynamic partial order reduction} (DPOR)
techniques are introduced (see, e.g., \cite{FG05,AAJS17}).
Intuitively speaking, DPOR techniques aim at producing only one
interleaving per Mazurkiewicz trace. 
This task is more natural in our context since we deal with traces
(that represent all derivations which are causal-consistent)
and logs (that represents all derivations which are
observationally equivalent). Therefore, there is no need to
consider DPOR techniques in our approach.
Concuerror \cite{CGS13} implements stateless model checking
for Erlang, and has been recently extended to also consider
\emph{observational equivalence} \cite{AJLS18}, thus achieving
a similar result as our technique regarding message races
despite the fact that the techniques are rather different.\footnote{To
the best of our knowledge, neither the semantics of Erlang 
nor the happened-before relation have been formalized in the
Concuerror approach.}
A key difference, though, is that Concuerror performs an 
extra pass on the schedulings, annotating
each message delivery with the patterns of the receive statement
\cite{AJLS18}. In this way, messages that do not match a given 
receive can be excluded from the computation of message races. 
This approach would involve adding message values and receive 
constraints to traces, which contrasts with our more lightweight
approach based on computing \emph{potential} sets of
message races (which, nevertheless, suffices in our particular
application within reversible debugging).

Another, related approach is the detection of race 
conditions for Erlang programs presented in \cite{CS10}. 
However, the author focuses on data races (that may occur when
using some shared-memory built-in operators of the language) 
rather than message races.
Moreover, the detection is based on a \emph{static} analysis, while
we consider a \emph{dynamic} approach to computing message
races.

Finally, regarding our tracing semantics 
(cf.\ Section~\ref{sec:applications}), 
it shares many similarities
with the logging semantics introduced in \cite{LPV19,LPV21}.
However, the logging semantics abstracts away process
mailboxes and message delivery since they are not needed
to replay an execution. Our semantics can be seen as a refinement
of the logging semantics
so that it is now closer to the actual semantics of the language.
%(indeed, we use the tracing semantics to produce a trace rather
%than a log).
As argued in Section~\ref{sec:motivation}, dealing explicitly 
with message delivery is essential in our
context in order to produce traces that can be used to 
compute message races.

%%%%%%%%%%%%%%%%%%%%%%%%%%%%%%%%%%%%%%%%%%%%%%%%%%%%%%%%%%
\section{Conclusions and Future Work}\label{sec:concl} 

In this paper, we have introduced a lightweight 
formalism to represent
concurrent executions in a message-passing language with
selective receives. In particular, we distinguish the notion of
trace from that of a log, which is essentially a 
representation of a Mazurkiewicz trace. 
Despite the simplicity of traces,
they contain enough information to analyze some common
error symptoms: blocked processes, lost or
delayed messages, and orphan messages.
Moreover, they can also be useful to compute 
(potential) message races, which can then be used
within our replay debugger \cauder\ \cite{cauder}
to explore alternative execution paths.
For this purpose, 
we have shown the details of a particular application
of our ideas in the context of causal-consistent 
reversible (replay) debugging of Erlang programs.

As future work, we plan to work on the complete formalization
of the extended framework for causal-consistent 
reversible replay debugging presented in 
Section~\ref{sec:cauder-races}, following a similar
approach as in \cite{LPV19,LPV21}.
%
%Moreover, we will also consider 
%the extension of the notion
%of trace in order to deal with some shared-memory operations
%(the case of several Erlang's built-in operators \cite{AS17})
%as well as to be able to check other types of properties
%(e.g., some user-defined properties).
%Finally, we also consider
%the extension of traces (and the tracing semantics)
%in order to deal with I/O operations, since this is essential to
%be able to work with a realistic language. 

%\subsubsection*{Acknowledgements.}
%
%The authors would like to thank Ivan Lanese for his useful remarks that 
%helped us to improve the new version of the CauDEr debugger.

%We would like to thank the anonymous reviewers and the participants of
%LOPSTR 2016 for their suggestions to improve this paper.

\bibliographystyle{splncs04}
\bibliography{biblio}

\begin{thebibliography}{10}
\providecommand{\url}[1]{\texttt{#1}}
\providecommand{\urlprefix}{URL }
\providecommand{\doi}[1]{https://doi.org/#1}

\bibitem{AAJS17}
Abdulla, P.A., Aronis, S., Jonsson, B., Sagonas, K.: Source sets: {A}
  foundation for optimal dynamic partial order reduction. J. {ACM}
  \textbf{64}(4),  25:1--25:49 (2017). \doi{10.1145/3073408}

\bibitem{AJLS18}
Aronis, S., Jonsson, B., L{\aa}ng, M., Sagonas, K.: Optimal dynamic partial
  order reduction with observers. In: Beyer, D., Huisman, M. (eds.) Proceedings
  of the 24th International Conference on Tools and Algorithms for the
  Construction and Analysis of Systems ({TACAS} 2018). Lecture Notes in
  Computer Science, vol. 10806, pp. 229--248. Springer (2018).
  \doi{10.1007/978-3-319-89963-3\_14}

\bibitem{CGS13}
Christakis, M., Gotovos, A., Sagonas, K.: {Systematic Testing for Detecting
  Concurrency Errors in Erlang Programs}. In: Proceedings of the 6th {IEEE}
  International Conference on Software Testing, Verification and Validation
  ({ICST} 2013). pp. 154--163. {IEEE} Computer Society (2013).
  \doi{10.1109/ICST.2013.50}

\bibitem{CS10}
Christakis, M., Sagonas, K.: {Static Detection of Race Conditions in Erlang}.
  In: Carro, M., Pe{\~{n}}a, R. (eds.) Proc. of the International Symposium on
  Practical Aspects of Declarative Languages (PADL 2010). pp. 119--133.
  Springer (2010)

\bibitem{erlang}
{Erlang} website. URL: \url{https://www.erlang.org/} (2021)

\bibitem{cauder}
Fabbretti, G., Gonz{\'{a}}lez{-}Abril, J.J., Lanese, I., Nishida, N., Palacios,
  A., Vidal, G.: {CauDEr} website. URL: \url{https://github.com/mistupv/cauder}
  (2021)

\bibitem{FG05}
Flanagan, C., Godefroid, P.: Dynamic partial-order reduction for model checking
  software. In: Palsberg, J., Abadi, M. (eds.) Proceedings of the 32nd {ACM}
  {SIGPLAN-SIGACT} Symposium on Principles of Programming Languages ({POPL}
  2005). pp. 110--121. {ACM} (2005). \doi{10.1145/1040305.1040315}

\bibitem{GLM14}
Giachino, E., Lanese, I., Mezzina, C.A.: Causal-consistent reversible
  debugging. In: Gnesi, S., Rensink, A. (eds.) Proceedings of the 17th
  International Conference on Fundamental Approaches to Software Engineering
  ({FASE} 2014). Lecture Notes in Computer Science, vol.~8411, pp. 370--384.
  Springer (2014)

\bibitem{God97}
Godefroid, P.: Model checking for programming languages using verisoft. In:
  POPL. pp. 174--186 (1997). \doi{10.1145/263699.263717}

\bibitem{GV21}
Gonz{\'{a}}lez{-}Abril, J.J., Vidal, G.: {Causal-Consistent Reversible
  Debugging: Improving CauDEr}. In: Morales, J.F., Orchard, D.A. (eds.)
  Proceedings of the 23rd International Symposium on Practical Aspects of
  Declarative Languages ({PADL} 2021). Lecture Notes in Computer Science, vol.
  12548, pp. 145--160. Springer (2021). \doi{10.1007/978-3-030-67438-0\_9}

\bibitem{GV21lopstr}
Gonz{\'{a}}lez{-}Abril, J.J., Vidal, G.: A program instrumentation for
  prefix-based tracing in message-passing concurrency. In: Angelis, E.D.,
  Vanhoof, W. (eds.) Proceedings of the 31st International Symposium on
  Logic-based Program Synthesis and Transformation ({LOPSTR} 2021) (2021),
  \url{http://personales.upv.es/gvidal/german/lopstr21/paper.pdf}

\bibitem{Lam78}
Lamport, L.: Time, clocks, and the ordering of events in a distributed system.
  Commun.\ {ACM}  \textbf{21}(7),  558--565 (1978). \doi{10.1145/359545.359563}

\bibitem{LNPV18}
Lanese, I., Nishida, N., Palacios, A., Vidal, G.: {CauDEr: A Causal-Consistent
  Reversible Debugger for Erlang}. In: Gallagher, J.P., Sulzmann, M. (eds.)
  Proceedings of the 14th International Symposium on Functional and Logic
  Programming ({FLOPS}'18). Lecture Notes in Computer Science, vol. 10818, pp.
  247--263. Springer (2018). \doi{10.1007/978-3-319-90686-7\_16}

\bibitem{LNPV18jlamp}
Lanese, I., Nishida, N., Palacios, A., Vidal, G.: A theory of reversibility for
  {E}rlang. Journal of Logical and Algebraic Methods in Programming
  \textbf{100},  71--97 (2018). \doi{10.1016/j.jlamp.2018.06.004}

\bibitem{LPV19}
Lanese, I., Palacios, A., Vidal, G.: Causal-consistent replay debugging for
  message passing programs. In: P{\'{e}}rez, J.A., Yoshida, N. (eds.)
  Proceedings of the 39th {IFIP} {WG} 6.1 International Conference on Formal
  Techniques for Distributed Objects, Components, and Systems ({FORTE} 2019).
  Lecture Notes in Computer Science, vol. 11535, pp. 167--184. Springer (2019).
  \doi{10.1007/978-3-030-21759-4\_10}

\bibitem{LPV21}
Lanese, I., Palacios, A., Vidal, G.: Causal-consistent replay reversible
  semantics for message passing concurrent programs. Fundam. Informaticae
  \textbf{178}(3),  229--266 (2021). \doi{10.3233/FI-2021-2005}

\bibitem{LC06}
Lei, Y., Carver, R.H.: Reachability testing of concurrent programs. {IEEE}
  Trans. Software Eng.  \textbf{32}(6),  382--403 (2006).
  \doi{10.1109/TSE.2006.56}

\bibitem{Maz86}
Mazurkiewicz, A.W.: Trace theory. In: Brauer, W., Reisig, W., Rozenberg, G.
  (eds.) Petri Nets: Central Models and Their Properties, Advances in Petri
  Nets 1986, Part II, Proceedings of an Advanced Course, 1986. Lecture Notes in
  Computer Science, vol.~255, pp. 279--324. Springer (1987).
  \doi{10.1007/3-540-17906-2\_30}

\bibitem{NPV16b}
Nishida, N., Palacios, A., Vidal, G.: A reversible semantics for {E}rlang. In:
  Hermenegildo, M., L\'opez-Garc\'{\i}a, P. (eds.) Proceedings of the 26th
  International Symposium on Logic-Based Program Synthesis and Transformation
  (LOPSTR 2016). Lecture Notes in Computer Science, vol. 10184, pp. 259--274.
  Springer (2017). \doi{10.1007/978-3-319-63139-4\_15}

\bibitem{SFB10}
Svensson, H., Fredlund, L.A., Earle, C.B.: {A unified semantics for future
  Erlang}. In: 9th ACM SIGPLAN workshop on Erlang. pp. 23--32. ACM (2010).
  \doi{10.1145/1863509.1863514}

\bibitem{Vid22submitted}
Vidal, G.: {Computing Race Variants in Message-Passing Concurrent Programming
  with Selective Receives} (2022), submitted for publication. Available from
  \url{http://personales.upv.es/gvidal/german/submitted/races.pdf}

\end{thebibliography}

\end{document}